\documentclass[aps,prb,twocolumn,showpacs,floatfix,amsmath]{revtex4}
\usepackage{amssymb}
\usepackage{float}

\usepackage{graphicx}

\begin{document}
\title{\textbf{Electrons and phonons in the ternary alloy CaAl$_{2-x}$Si$_x$} as a function of composition}

\author{Matteo Giantomassi,$^{1,2}$}

\altaffiliation[Present Address: ]{ Unit\'e de Physico-Chimie et de Physique des Mat\'eriaux, Universit\'e Catholique de Louvain, 1 place Croix du Sud, B-1348 Louvain-la-Neuve, Belgium}
\author{ Lilia Boeri,$^3$ and
Giovanni B. Bachelet$^1$}

\affiliation{$^1$INFM Center for Statistical Mechanics and Complexity and Dipartimento di Fisica, Universit\`a di Roma ``La Sapienza'', Piazzale A. Moro 2, I-00185 Roma, Italy \\
$^2$ Istituto dei Sistemi Complessi, CNR, Via dei Taurini 19, 00185
  Roma, Italy \\
$^3$Max-Planck-Institut f\"{u}r Festk\"{o}rperforschung,
 Heisenbergstr.~\!\!1, D-70569 Stuttgart, Germany}
\date{\today}
\begin{abstract}
We report a detailed first-principles study of the structural,
electronic and 
vibrational properties of the superconducting C$_{32}$ phase of the 
ternary alloy CaAl$_{2-x}$Si$_x$, both in the experimental range $0.6 \leq x \leq 1.2$, for which the alloy has been synthesised, and in the theoretical limits of high aluminium and high silicon concentration.
Our results indicate that, in the experimental range, the dependence
of the electronic bands on composition is well described by a rigid-band
model, which  breaks down outside this range. Such a breakdown, 
in the (theoretical) limit of high aluminium concentration, is connected to the appearance of vibrational 
instabilities, and results in important differences between CaAl$_2$ and MgB$_2$.
Unlike MgB$_2$, the interlayer band and the out-of-plane phonons play a major
role on the stability and superconductivity of CaAlSi and related C$_{32}$
intermetallic compounds.
\end{abstract}

\maketitle
\section*{INTRODUCTION}
The discovery~\cite{mgb2discovery} of superconductivity at T$_{c}\!=\!39$ K in  magnesium diboride (MgB$_2$) 
is remarkable not only for its high critical temperature relative to conventional superconductors, 
but also because it has destroyed prejudices and created new hopes on electron-phonon (e-ph) superconductors.
MgB$_2$ crystallises in the AlB$_2$-type structure, also called  C$_{32}$ phase 
($P6/mmm$, No.191),~\cite{c32type} forming hexagonal layers of graphene-like sheets of B atoms intercalated with Mg planes;
the synthesis and study of materials with the AlB$_2$ structure has therefore
attracted a great deal of interest in the last few years.

In the present paper we will focus on the ternary alloy
CaAl$_{2-x}$Si$_x$, which has been recently synthesised~\cite{CaAlSiordered} for a relatively
wide range of composition $0.6 \leq x \leq 1.2$ and, in this range, has been reported to be 
isostructural to MgB$_2$, with honeycomb, graphene-like sheets of randomly distributed silicon and aluminium atoms, and alkaline-earth atoms intercalated between 
these sheets.~\cite{CaAlSiordered}
Superconductivity has been detected in the whole composition range; the critical temperature T$_c(x)$ strongly depends on composition, showing a sharp peak at $x\!=\!1$ (where T$_c\!=\!7.8$ K), and dropping off rapidly as the Al/Si ratio departs from unity.~\cite{CaAlSiordered}

Several experiments\cite{CaAlSi_pressione_exp,unconventional_SC_CaAlSi,CaAlSi-Fermi} and  {\em ab initio} calculations~\cite{SheinCaAlSi,MazinCaAlSi,CaAlSifononi-cinesi,CaAlSi_pressure_effects} have concentrated on the $x\!=\!1$ case (i.e. CaAlSi).
Concerning the superconducting mechanism,  Mazin \emph{et al.}~\cite{MazinCaAlSi} have suggested that the available experimental data
  could be reconciled with a traditional e-ph scenario, assuming the existence
  of a soft phonon mode ($\sim$ 30-40 cm$^{-1}$); recently 
Huang \emph{et al.}~\cite{CaAlSifononi-cinesi,CaAlSi_pressure_effects} have
  calculated the phonon  spectrum and electron-phonon coupling of CaAlSi, revealing the existence of a very soft (and, at some wavevector, even unstable) 
$B_{1g}$ branch, which plays an important role in its
  superconducting properties; they also claim that this mode, which
  gets softer under pressure, can be invoked to explain the positive pressure effect on T$_c$.~\cite{CaAlSi_pressure_effects}

In contrast to the $x\!=\!1$ case, neither the wide stability range $ 0.6 \leq x \leq 1.2 $ of  of the C$_{32}$ phase, nor its instability outside this range (the limit of high aluminium or silicon concentration), nor any other physical property of CaAl$_{2-x}$Si$_x$ has been, so far, theoretically studied as a function of composition, at least to our knowledge.

Our paper reports the first detailed \emph{ab initio} analysis of the
structural, electronic and vibrational properties of this ternary alloy, both in the composition range for which it has been actually synthesised, and outside it, in the theoretical limits of high aluminium and high silicon concentration. The former limit ($x\!\rightarrow\!0$) is of particular interest, since Ca and Al have the same valence configuration as Mg and B, respectively, and, therefore, the hypothetical\cite{FN1}
C$_{32}$ phase of CaAl$_2$  would not only be isostructural, but also isoelectronic to MgB$_2$, attracting attention as a possible new superconductor or, at least, as a relevant contribution to the understanding of MgB$_2$ and related superconducting alloys.\cite{MGBC}

The work is organised as follows. In Sec.~\ref{sect:computational_details} we briefly describe the computational details of our study, including the virtual crystal approximation
adopted for a disordered alloy.
In Sec.~\ref{sect:electrons_x=1} we discuss the electronic bands of the $x\!\!=\!\!1$ case, considering both an 
ordered phase with in-plane Al-Si ordering and the disordered alloy, and compare our results with previous calculations\cite{MazinCaAlSi,CaAlSifononi-cinesi} and ARPES data.~\cite{CaAlSi-Fermi}  We also introduce a third, fictitious compound  $\Box^{2+}$AlSi, with calcium ions replaced by a
uniform background of positive charge, whose electronic bands will help us identify the role of calcium ions.  
The vibrational properties and the electron-phonon coupling in CaAlSi and
$\Box^{2+}$AlSi are presented in Sec.~\ref{sect:e-ph} while, 
in Sec.~\ref{sect:properties_doped}, 
we describe how the different Al/Si content modifies the electronic and
structural properties of CaAl$_{2-x}$Si$_x$.  
  
\section{Computational details} \label{sect:computational_details}

The electronic, structural and vibrational properties of all our crystals were calculated in the framework of the density functional theory\cite{DFT:KS,DFT:HK} and of the
density functional perturbation theory\cite{DFT:BaroniRMP},
with Kohn-Sham wavefunctions expanded on a plane-wave basis limited by a 40 Ryd cutoff energy.~\cite{ABINIT,PWscf}
We used Troullier-Martins~\cite{Troullier-Martins} norm-conserving pseudopotentials generated by the FHI98PP package~\cite{fhi98pp} and consistently
employed the GGA-PBE functional~\cite{PBEfunctional} to approximate the exchange-correlation energy functional.
The {\bf k}-space integration (electrons) was approximated with a $12\times12\times12$ Monkhorst-Pack grid~\cite{MPgrid} for the self-consistent cycles; 
with the more accurate tetrahedron method~\cite{tetrahedron} and a $20\times20\times20$ mesh 
for the electronic density of states (DOS); a $18\times18\times18$ Monkhorst-Pack grid for the electron-phonon coupling.
After setting the smearing parameter~\cite{Coldsmearing} to 0.03 Ryd, the electronic total
energies were converged to 0.1 mRy, while the phonon frequencies were converged  within 5 cm$^{-1}$ with respect to the k-point sampling; we checked that a different choice of the smearing parameter between 0.02 Ryd and 0.04 Ryd would not change our frequencies by more than 5 cm$^{-1}$. 
Dynamical matrices were evaluated for phonon wave vectors {\bf q} on a
$6\times6\times6$  grid, from which the phonon dispersion
was obtained by Fourier interpolation.

To describe a disordered CaAl$_{2-x}$Si$_x$ alloy we employed the Virtual Crystal Approximation (VCA)~\cite{VCA}, 
thus considering a $P6/mmm$ structure where the sites of the hexagonal graphene sheets are occupied 
by an alchemical element whose ionic pseudopotential is defined as 
$\hat{v}_{\text{ion}}^\text{alch}=({1-x}/{2})\hat{v}_{\text{ion}}^\text{Al}+({x}/{2})\hat{v}_{\text{ion}}^\text{Si}$. 
The use of the VCA permits us to investigate the whole composition range $x \in \left[0,2\right]$ without introducing superstructures (which on the other hand were not experimentally
revealed in the samples examined with a powder $x-$ray diffraction technique).~\cite{CaAlSiordered}
The VCA should be justified in our case, since Al and
Si are neighbours in the periodic table and have a very similar
electronic structure.

\section{Electrons in ${\rm CaAlSi}$, ${\rm Ca(AlSi)}$ and \,$\Box^{2+}{\rm\!\!AlSi}$} \label{sect:electrons_x=1}

The $x\!\!=\!\!1$ case requires a thorough discussion,
since this is the composition to which most experimental and theoretical results refer. Furthermore, there is still an open question concerning the
actual crystal structure, whether a chemically disordered phase  ( where each site of the hexagonal layers 
is randomly occupied by either Al or Si, with equal probability), or a crystal
structure with some Al/Si ordering. 
The sharp $x$-ray diffraction patterns  and the relatively high phase purity measured in CaAlSi (as opposed to the case of
CaAl$_{2-x}$Si$_x$ with $x\!\neq\!1$) 
suggest a possible Al/Si ordering for $x\!=\!1$.~\cite{CaAlSiordered}
 This is also what Mazin \emph{et al.}~\cite{MazinCaAlSi} suggest, proposing a crystalline structure with atomic in-plane ordering 
(i.e., graphene-like sheets with alternated Al and Si atoms) of
$P\overline{6}m2$ (No.~187) symmetry,~\cite{c32type} obviously
lower than the C$_{32}$ symmetry described above,
and apparently favourable with respect to other possible
ordering patterns.

An analysis of powder $x$-ray diffraction patterns performed by Imai~\emph{et al.}~\cite{CaAlSidisorderd} suggests, instead, 
the disordered phase.  
The question is still under debate and no conclusive experimental measure has been reported yet. 

To elucidate this controversy we have repeated Mazin's first-principles
calculations, 
considering both the ordered $P\overline{6}m2$ structure and the 
disordered $P6/mmm$ VCA phase.
Our comparison of these results with the ARPES experiments~\cite{CaAlSi-Fermi}
gives additional arguments in favour of the ordered phase.
Moreover, the VCA phase represents a crucial reference for our subsequent 
rigid-band model.
Hereafter we will indicate our disordered  VCA phase as Ca(AlSi) and the ordered phase as CaAlSi.
The CaAlSi and Ca(AlSi) lattice parameters were obtained by optimizing the
structural parameters $c$ and $a$ up to a maximum residual stress of 1 MPa;
the numerical values are listed in Table~\ref{table:lattice_parameters}.

\begin{table}[h!]
\caption{Experimental lattice parameters at NTP\cite{CaAlSiordered}
and optimised lattice parameters for the ordered CaAlSi and the VCA-disordered Ca(AlSi) obtained in this work, based on GGA-PBE exchange-correlation and pseudopotentials.}
\begin{ruledtabular}
\begin{tabular}{ccccc}
     & $a$({\em a.u.})   & $c$ ({\em a.u.})   & $c/a$ &
\\[1ex] 
Exp      & 7.916 & 8.315 & 1.050 & 
\\[1ex] 
\hline
\\[1ex] 
CaAlSi   & 7.979 & 8.446 & 1.058 & 
\\[1ex] 
\% error    & 0.79 & 1.57  & 0.76  
\\[1ex] 
\hline
\\[1ex] 
Ca(AlSi) & 7.965 & 8.485 & 1.065 & 
\\[1ex] 
\% error     & 0.62 & 2.04 & 1.42  &      
\\[1ex] 
\end{tabular}
\end{ruledtabular}
\label{table:lattice_parameters}
\end{table}

The agreement of the calculated lattice parameters with the experiment is good (slightly better for the ordered phase),
with errors of the order of 1$\%$. 
The corresponding electronic bands 
are shown in Fig.~\ref{fig:bands_for_x=1} for the ordered structure (upper
panel) and the disordered VCA phase (middle panel); they 
 agree quite well with the previous
 calculations,~\cite{SheinCaAlSi,CaAlSifononi-cinesi}, and, in particular,
 with the corresponding full-potential results ~\cite{MazinCaAlSi}.

We will start our discussion with the energy bands of the ordered CaAlSi structure. 

\subsubsection*{CaAlSi}

In CaAlSi the  $sp^2$-hybrids of aluminium and silicon form three bonding $\sigma$ bands:
the lowest occupied $\sigma$ band is formed by $s$-like states and is  
separated by a gap from the other two $\sigma(2p_{x,y})$  bands, which are mainly formed by the $2p_{x}$, $2p_{y}$ states of Al and Si.
The $\sigma(2p_{x,y})$ bands have a quasi two-dimensional (2D) character, with a small dispersion along $k_z$ (the $\Gamma$-A direction) very similar 
to what is  observed in MgB$_2$.\cite{mg:kortus1,mg:pickett} We will refer to  the $\sigma(2p_{x,y})$ bands as $\sigma$ bands in what follows.
The $2p_{z}$ states of Al and Si form a bonding ($\pi$) and an anti-bonding ($\pi^*$) band, which are completely three-dimensional (3D).

Because of the reduced crystal symmetry, the $\pi$ and $\pi^*$ bands do not cross at $K$, but
are separated by a gap in the spectrum.
And, unlike MgB$_2$, both the bonding $\pi$ and the bonding $\sigma$ bands are fully occupied, because here there is one more electron per unit cell; so these bands cannot play any role for superconductivity.
Moreover, because of the low $c/a$ ratio, the $\pi$ and $\pi^*$ bands acquire a substantial
dispersion along $k_z$, much higher than in graphite and MgB$_2$;
the final result is that,
  in the ``upper half" of the
Brillouin zone ($k_z\simeq \pm\pi/c$, $A-L-H-A$ path), a good fraction of the $\pi^*$ band is occupied too, and the antibondig $\pi^*$ band,
of 3D character, crosses the Fermi level.

Then, as we see in the upper panel of Fig.~\ref{fig:bands_for_x=1}, a second
 band crosses the Fermi level. It's a  3D band whose wavefunctions extend in
 the interlayer region. 
Its bottom has a prevailing Ca\,$3d_{z^2 - r^2}$
 character, yet the band as a whole may be easily traced back to the so-called
 graphite {\em interlayer band};~\cite{posternack} 
therefore, in what follows, we will stick to the name.
This band is empty in graphite
 and  MgB$_2$,  but here, in CaAlSi, it's partially full, because of the
 different electron count.

As we see in the upper panel of Fig.~\ref{fig:Fermisurface}, the CaAlSi Fermi
surface resulting from 
these two bands has two topologically disconnected sheets. 
The first, outer sheet is a sixfold gear-like surface: 
a wide empty cylinder, whose axis is along $k_z$, connected to  
the neighbouring  Brillouin zones in the $k_{xy}$ plane along the $\Gamma$-M and equivalent directions  
by six sleeves, stretching out from its side surface.
This sheet of the Fermi surface actually derives from both
  bands: near $k_z$=0 its character is mainly interlayer; 
near k$_z$=$\pm\pi/c$  it is mainly $\pi^*$.
   The second, inner sheet is a hexagonal lens-like structure centered at
   $\Gamma$. It exclusively arises from the interlayer band, which cuts
   $\epsilon_F$ (upper panel of Fig.~\ref{fig:bands_for_x=1})
 around the centre of the Brillouin zone (BZ).
This structure is also observed in the ARPES data,~\cite{CaAlSi-Fermi} which
 have revealed the presence of occupied states around $\Gamma$ forming
a hexagonal structure.

\subsubsection*{Ca(AlSi)}

Comparing the Ca(AlSi) and CaAlSi bands of Fig.~\ref{fig:bands_for_x=1}, we
can notice that, because of the increased symmetry, the $\pi-\pi^*$ energy gap
disappears in the VCA: now the $\pi$ bands cross at K, forming an electronic
pocket at E$_F$.
 
The most important difference is that the inner sheet of the Fermi surface, a
large hexagonal lens in the ordered CaAlSi, shrinks to almost nothing in the
VCA  (a tiny circular lens, see lower panel of Fig.~\ref{fig:Fermisurface}),
because now the Fermi level
cuts the interlayer band almost at its bottom, at $\Gamma$ 
(see middle panel of Fig.~\ref{fig:bands_for_x=1}).
Otherwise, the outer sheet is almost identical, differing
from  the corresponding one in CaAlSi only near the K point because of the additional small electronic pocket.

The lack of a sizable hexagonal lens-like structure makes the Fermi surface of
the disordered structure less compatible with the experimental ARPES map.
Furthermore, unlike the ordered phase, the band structure of the disordered Ca(AlSi) (middle panel of Fig.~\ref{fig:bands_for_x=1}) cannot explain~\cite{mythesis}  an ARPES high-intensity 
region located at $\sim -0.8$ eV, derived from $\pi$ states, which is clearly visible in Fig. 2 of Ref.~\onlinecite{CaAlSi-Fermi}. Our results seem thus to confirm the existence of in-plane ordering in the compound synthesised with equal amounts of Al and Si.

While the bands of the ordered phase appear in better agreement with the ARPES experiment at $x\!\!=\!\!1$, the VCA band structure of the disordered Ca(AlSi),
middle panel of Fig.~\ref{fig:bands_for_x=1}, represents a very convenient reference for a rigid-band description of CaAl$_{2-x}$Si$_x$
at $x\!\neq\! 1$, a model which will prove quite accurate in the experimental range of compositions $x \in
\left[0.6;1.2\right]$.
In a rigid-band framework we find it useful to report, on the right side of Fig.~\ref{fig:bands_for_x=1},
the Integrated Density of States (IDOS), with black dashed lines corresponding
to the band edges of the interlayer, $\sigma$, and $\pi$ bands.

In this way we may see that, below
8.8 electrons, the $\pi$ bonding band in the $k_z=0$ begins to be emptied;
below 8.5 electrons, the interlayer band is completely empty, and
below 8.25 electrons the $\sigma$ bonding band in the $k_z=\pm\pi/c$
plane begins to be emptied (see Section \ref{sect:properties_doped}).

\subsubsection*{$\Box^{2+}$AlSi}

\begin{figure}[htb!]
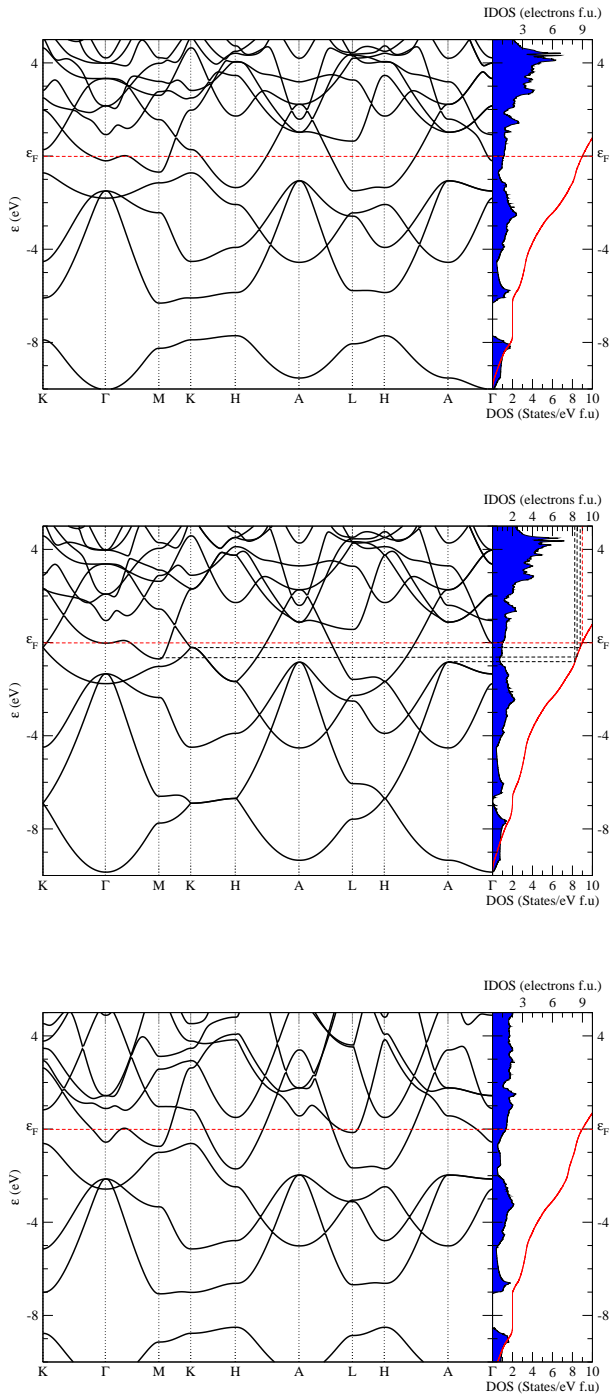

\begin{center}
\includegraphics[width=8cm]{1_0}
\end{center}
\vspace{0.2cm}
\begin{center}
\includegraphics[width=8cm]{1_0vc}
\end{center}
\vspace{0.2cm}
\begin{center}
\includegraphics[width=8cm]{alsi}
\end{center}
\caption{Upper panel: energy bands (black), DOS (blue) and integrated DOS (IDOS, red) for CaAlSi. 
Middle panel: same as above for the disordered Ca(AlSi); the 
three black dashed lines pinpoint, along the IDOS, the number of electrons  at which, in a rigid band scheme, a few relevant bands begin to be emptied or filled (see text).
Lower panel: same as the upper panel, for the fictitious $\Box^{2+}$AlSi.}

\label{fig:bands_for_x=1}
\end{figure}

\begin{figure}[!hbt]
\begin{center}
\includegraphics[width=3.5cm, height=3.5cm,keepaspectratio]{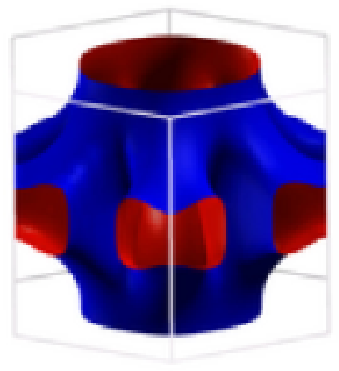}
\includegraphics[width=3.5cm, height=3.5cm,keepaspectratio]{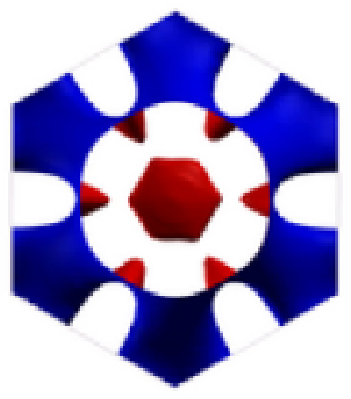}
\end{center}
\begin{center}
\includegraphics[width=3.5cm, height=3.5cm,keepaspectratio]{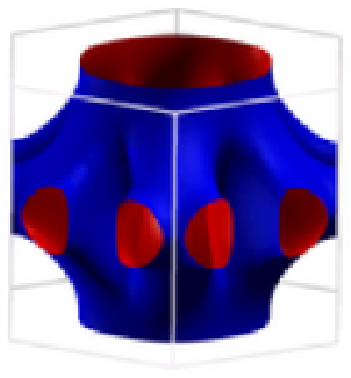}
\includegraphics[width=3.5cm, height=3.5cm,keepaspectratio]{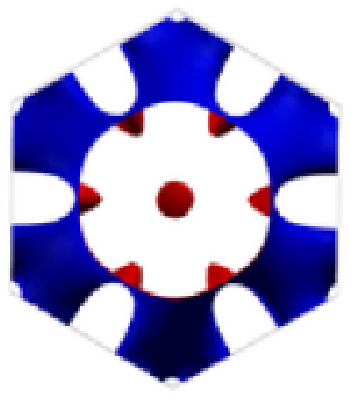}
\end{center}
\caption{Fermi surface and Brillouin zone of CaAlSi in the in-plane ordered (upper panels) and disordered (lower panels) structure. The outer sheet originates from both the $\pi^*$ (around A) and the interlayer band (around $\Gamma$). The inner sheet (a lens-like structure) exclusively originates from the interlayer band at $\Gamma$, and is much larger, with more evident hexagonal symmetry, in the ordered structure (upper panel). We have chosen the blue color for the outer surface of the  gear-like structure, and the red color for its inner surface. The surface of the inner, lens-like structure, is also red. With this choice the top view (right panels) happens to show the $\pi^*$ part of the Fermi surface in blue and the interlayer part in red.} 
\label{fig:Fermisurface}
\end{figure}

If a few differences in the electronic bands have helped us discriminate
between the ordered and disordered phase, their similarity is overwhelming;
the band structures in the upper and middle panels of
Fig.~\ref{fig:bands_for_x=1} are, in fact, almost identical. Now we want to
point out that the same striking similarity extends to the lower panel, where we show the bands of $\Box^{2+}$AlSi. This fictitious compound has the same number of electrons and the same lattice parameters as CaAlSi, but the Ca$^{2+}$ ion has been replaced by a homogeneous positive background $\Box^{2+}$. 

We see that the absence of the calcium ion 
doesn't dramatically change the band dispersion at $\epsilon_F$: the $\pi^{\star}$ is 
unchanged, while the interlayer band shows slight modifications. This indirectly confirms the interlayer character of this band. 
The important differences with respect to CaAlSi are located above $\epsilon_F$, where the bunch of  
empty Ca-$d$ bands (and the corresponding large hump in the DOS) is clearly (and obviously) absent.

\section{Phonons and e-ph coupling in ${\rm CaAlSi}$ ($x=1$)} \label{sect:e-ph}

\begin{figure}[h!tbp]
\begin{center}
\includegraphics[width=8cm]{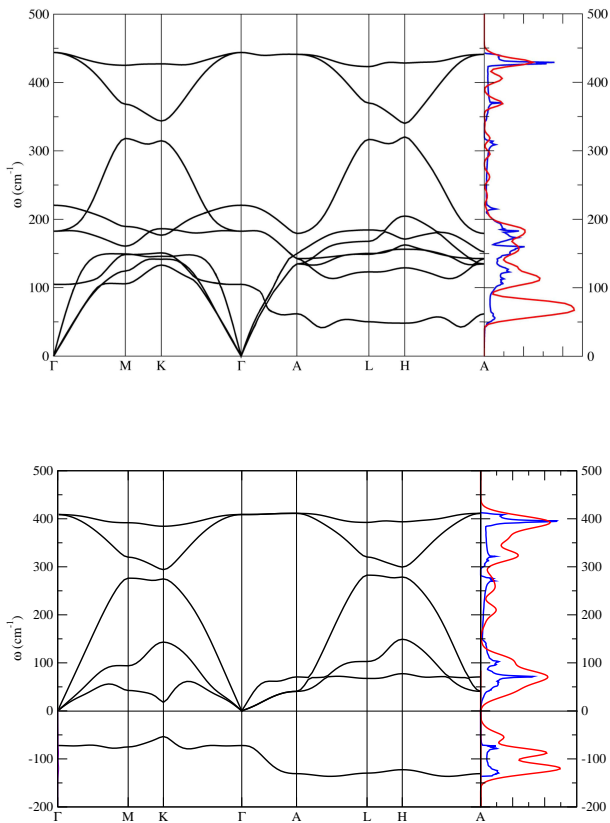}
\end{center}
\vspace{0.5cm}
\begin{center}
\includegraphics[width=8cm]{ph_alsi}
\end{center}
\caption{Upper panel: Phonon dispersion, phonon DOS (blue) and Eliashberg
  function (red) $\alpha^2F(\omega)$ for CaAlSi. Notice that the Eliashberg
  function has a large peak around 50 cm$^{-1}$, corresponding to the soft
  B$_{1g}$ branch.
Lower panel: same as above but in the fictitious $\Box^{2+}$AlSi;
the B$_{1g}$ branch has an equally large e-ph coupling, but now its frequency
  are imaginary (negative in the plot)  in the entire BZ, revealing a
  dynamical instability (see text).}
\label{fig:phonons}
\end{figure}

Having become convinced that an atomic in-plane ordering takes place for $x\!=\!1$,
we will discuss the phonon dispersion only in the ordered phase, shown in the upper panel of Fig.~\ref{fig:phonons}. 
At the $\Gamma$ point, in addition to the acoustical modes, there are four distinct optical modes:  one doubly degenerate E$_{1u}$ mode 
(vibrations of Ca atoms against the Al-Si planes along  $x,y$), one A$_{2u}$ mode (Ca vibrations against Al-Si planes along $z$), one doubly degenerate E$_{2g}$ mode (in-plane Al-Si bond-stretching mode), and one B$_{1g}$ mode (Al and Si displaced along $z$ in opposite directions).  
Table~\ref{table:ph_frequencies}  compares the frequencies $\omega$ of the optical modes at $\Gamma$ obtained  in this work 
with those reported in previous full-potential calculations.
For comparison, in the same table we also report the corresponding phonon
frequencies for MgB$_2$ \cite{mg:kong}. 

\begin{table}[h!]
\caption{Calculated frequencies (cm$^{-1}$) of the zone-center optical phonons in ordered CaAlSi, as obtained here (asterisk) and in previous calculations. MgB$_2$ is shown for comparison.} 
\begin{ruledtabular}
\begin{tabular}{cccccc}
& B$_{1g}(\Gamma)$ & E$_{1u}(\Gamma)$ & A$_{2u}(\Gamma)$ & E$_{2g}(\Gamma)$  &  
\\[1ex]
\\[1ex]
CaAlSi$^\star$&                      105& 183 & 221 & 444
\\[1ex]
CaAlSi \cite{CaAlSifononi-cinesi}&   99& 194& 222& 455
\\[1ex]
CaAlSi \cite{MazinCaAlSi}&          100& 187& 212& 456
\\[1ex]
\\[1ex]
MgB$_2$ \cite{mg:kong}& 692& 335& 401& 585
\\[1ex]
\end{tabular}
\end{ruledtabular}
\label{table:ph_frequencies}
\end{table}

The most striking feature is the presence of a very soft $B_{1g}$ mode. First intuition, inspired by the corresponding mode in MgB$_2$ and other diborides\cite{mg:kong,mg:satta}, would place this mode, which only involves vertical displacements of the lighter Al and Si atoms, at frequencies higher than those involving Ca too. We find it, instead, as the lowest-frequency optical mode not only at the $\Gamma$ point (Table~\ref{table:ph_frequencies}), but also in the entire ``upper half" of the BZ (Fig.~\ref{fig:phonons}). This finding can be reconciled with the mass argument only by a much weaker vertical force constant; and indeed, unlike the other compounds, CaAlSi has a partial filling of the $\pi^*$ bands in the ``upper half'' of the Brillouin zone
($k_z\simeq \pm\pi/c$, $A-L-H-A$ path), which weakens the bond in the vertical direction.

Our phonon frequencies at $\Gamma$ (Table~\ref{table:ph_frequencies}) are in
good agreement with all previous calculations; the same agreement is found
with the entire phonon spectrum obtained by Huang \emph{et al.}
\cite{CaAlSifononi-cinesi}, with a notable exception: our  $B_{1g}$ branch 
 (upper panel of Fig.~\ref{fig:phonons}) has a frequency of
$\sim$~50~cm$^{-1}$ and is almost dispersionless in the $q_z=\pm\pi/c$ plane;
Huang \emph{et al.} find, instead, imaginary frequencies along the A-L
path. We find no such dynamical instability either there, or, in fact, in any
other phonon branch and wavevector. Our results agree with the overwhelming
experimental evidence \cite{CaAlSiordered,CaAlSi-Fermi,LAVES1} in favour of the structural stability of
CaAlSi, and match the remarkable guess of Mazin {\em et al.},
 that, outside $\Gamma$, a soft mode must be there at about 35-40 cm$^{-1}$
for a consistent explanation of both normal and superconducting properties.

In the upper panel of Fig.~\ref{fig:phonons}, right side, the calculated
Eliashberg function $\alpha^2 F(\omega)$, shown in red, is compared to the
phonon DOS, shown in blue. The Eliashberg function evidently does not follow
the phonon DOS, and shows a pronounced peak at $\sim$ 50 cm$^{-1}$, clearly
arising from the (anomalously low) $B_{1g}$ branch. Electrons mainly couple
with this branch; the largest contribution to the integrated superconducting
parameter $\lambda$ will appear along the $\Gamma$-A-L path. 
This coupling is mainly interband in character: a B$_{1g}$ displacement mixes the interlayer and $\pi^*$ bands, while,
by symmetry, the corresponding $\pi$-$\pi$ coupling remains negligible
\cite{MAURI_GRAF,pippo3}.

Since, as discussed in Sec.~\ref{sect:electrons_x=1}, Fermi-level interlayer electrons are near $k_z$ = 0, while Fermi-level $\pi^*$
electrons are near $k_z=\pm\pi/c$, the electron-phonon coupling
happens at wavevectors  in the ``upper half" of the BZ
($q_z \simeq \pm\pi/c$, $A-L-H-A$ path).
The total $\lambda$, obtained as a weighted sum over individual phonon linewidths, is 0.66. The critical temperature can then be estimated from the Dynes equation
\begin{equation}
  \label{eq:Dynes}
 T_c=\frac{\omega_{log}}{1.2}\,
 \exp\left(-\frac{1.04(1+\lambda)}{\lambda-\mu^*(1-0.62 \lambda)}\right); 
\end{equation}
with a calculated logarithmic average frequency of 151~K, and
assuming $\mu^*=0.1$, we obtain a T$_c$ of 5.8~K, in reasonable
agreement with the experimental T$_c$ of 7.9~K.

In the lower panel of Fig.~\ref{fig:phonons} we report the same vibrational
properties (phonon dispersion, DOS and Eliashberg function) as in the upper
panel, but calculated for the fictitious system  $\Box^{2+}$AlSi. Here we
obviously have only 6 out of 9 phonon branches, since the Ca atoms are
missing. Let us focus on the three optical modes, which correspond to the 
two E$_{2g}$ branches (degenerate at $\Gamma$) and the B$_{1g}$ branch of the
original compound.
 Our results show that, while the two E$_{2g}$ modes are more or less
 unchanged,
 the substitution of Ca ions with a uniform background of positive 
charge drives the B$_{1g}$ branch towards imaginary frequencies (shown as negative in Fig.~\ref{fig:phonons}) across the entire Brillouin zone, indicating that our fictitious compound is not stable in this structure.

The Eliashberg function $\alpha^2 F(\omega)$ (red line) is thus well defined
for all branches except this one. 
At imaginary frequencies we may extend, as done here, its
definition by taking their modulus.
 Such an extended Eliashberg function cannot be integrated to yield a total
 $\lambda$, but may be used to illustrate the frequency distribution of the
 e-ph interaction,\cite{FN2}
which turns out to be very similar to the real CaAlSi (upper
panel). This is not surprising, because in $\Box^{2+}$AlSi the
relevant, gear-like structure of the Fermi surface (not shown) is very
similar to that of ordered and disordered CaAlSi (see Fig.~\ref{fig:bands_for_x=1}).
The only difference is that,
in the fictitious $\Box^{2+}$AlSi (lower
panel), the e-ph peak corresponding to the B$_{1g}$ branch, is shifted to
negative frequencies;  
we may therefore conclude that, although the Ca ion is not so important to
define the CaAlSi electronic bands at $\epsilon_F$, it plays a crucial role
for its structural stability, because, with respect to the equivalent
jellium-like charge, a positive ion placed between Al-Si layers is more effective
both in binding them and in hindering the Al and Si vertical displacements .

In summary, we find an entire phonon branch of imaginary frequencies
in the fictitious compound $\Box^{2+}$AlSi, but we do not find any imaginary 
frequency in the phonon spectrum of the actual (ordered) CaAlSi, in contrast
to previous, 
full-potential calculations. In CaAlSi we only find a very soft branch, which
suggests 
the proximity of a structural transition. This single phonon mode, 
which is soft in the ``upper half" of the Brillouin zone, dominates the electron-phonon coupling.

We suspect that a similar situation also holds at compositions other than
$x\!\!=\!\!1$, and, in the following section, we explore in some detail the
relation between lattice stability and band structure in the whole composition
range. As we will see, the results confirm our hint, at least in the limit of
high-aluminium concentration, 
and provide a useful reference and a strong motivation for a future, complete
study of the electron-phonon 
interaction as a function of $x$.

\section{Physical properties of ${\rm CaAl}_{\,2-x}{\rm Si}_x$} \label{sect:properties_doped}

In this section  we will give a quantitative description and attempt a
qualitative interpretation of how the composition $x$  affects the electronic
and structural properties of the C$_{32}$ phase of CaAl$_{2-x}$Si$_x$. 
We will consider a dense grid of $x$ values in the range $\left[0,2\right]$,
which includes both the physical 
$\left[0.6,1.2\right]$ and the unphysical range. 
For each  $x$-value considered, the hexagonal unit cell has been optimised up to a residual stress of 1 MPa,
and band structure calculations have been performed in the relaxed
configuration. For this dense grid of $x$ values a slightly lower cutoff of 
36 Ryd was used for the plane-wave basis.
In Fig.~\ref{fig:lattice_parameters_rigidband} we report, as a function of $x$,  the optimised lattice parameters (upper panel) and $g(\epsilon_F)$, the calculated DOS at the Fermi level (lower panel, black). We start our discussion from the experimental $x$-range.    

\subsubsection*{Experimental range, $0.6 \leq x \leq 1.2$}

Taking into account that we have modelled the disorder using the VCA, and that the experiments  detect several types 
of impurities, our theoretical in-plane bond length appears in reasonable agreement with the experimental data; the largest discrepancies are observed close to $x=0.6$, which is the lowest aluminium concentration for which the series has been synthesised. 
The somewhat worse agreement for $c$ is common to the ordered phase and may be traced back to a systematic error of the 
approximate exchange-correlation functional.

In the whole experimental range, $a$ monotonically decreases with $x$, while
$c$ is almost constant.

The small (less than 3$\%$) variations in the lattice parameters correspond to minor
modifications in the band structure: in the experimental $x$-range the calculated electronic structure of CaAl$_{2-x}$Si$_x$
(not shown) can be accurately described by a simple rigid-band model, where the Fermi level changes only according to the different number of electrons per cell $N_\text{e}(x)=8+x$, see Fig.\ref{fig:bands_for_x=1}.
 
An indirect proof of the validity of this model is given in the lower panel of 
Fig.~\ref{fig:lattice_parameters_rigidband} where
we plot, as a function of $x$, the self-consistent DOS at the Fermi level, 
$g(\epsilon_F)$, of CaAl$_{2-x}$Si$_x$ and the corresponding value 
for a rigid-band filling  $N_\text{e}(x)$ of the (fixed) bands of Ca(AlSi),
i.e., the bands shown in the middle panel of Fig.\ref{fig:bands_for_x=1}. 
As we can see, in the composition range $0.6 \leq x \leq 1.2$, where the
alloy is   
experimentally found to be stable in this crystalline structure, the two curves are almost identical;
deviations start to  appear at the extrema of this range. Then, outside this range, the rigid-band model completely breaks down. 
The dependence of $g(\epsilon_F)$ on composition thus correlates with rigid-band and structural 
properties. Its almost monotonic variation in the experimental range of stability $0.6 \leq x \leq 1.2$ does not, instead, correlate with 
the observed trend in the superconducting transition:
the experimental T$_c (x)$ has a sharp peak at $x=1$ and drops off rapidly as the Al/Si ratio is changed in either directions.~\cite{CaAlSiordered} 
This suggests, by exclusion,  that the experimental trend in T$_c (x)$ must be  associated with changes of the phonon frequencies 
and/or electron-phonon coupling strength as a function of composition.

\subsubsection*{High aluminium concentration, $x < 0.6$}

For $x < 0.6$ the $a$ lattice parameter still increases as $x$ decreases, its slope being larger than in the experimental range of stability. Much more peculiar is the  behaviour of the $c$  parameter. Reverting the trend observed in the experimental range of stability,  $c$ decreases as $x$ decreases, until it equals the $a$ value, for  $x\sim 0.3-0.4$; then, at $x=0.2$, it undergoes an abrupt change, such that the $c$/$a$ ratio jumps by $\sim 20 \%$.
Our lattice dynamical calculations show
that this change is accompanied by the occurrence of dynamical instabilities.

\begin{figure}[h!tbp]
\begin{center}
\vspace{0.6cm}
\includegraphics[width=8cm]{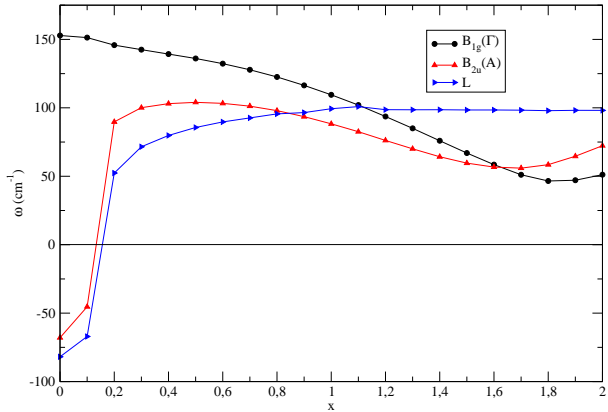}
\end{center}
\caption{Selected phonon frequencies of the spectrum of CaAl$_{2-x}$Si$_x$ as a
  function of the Silicon content $x$. The branch which corresponds to the
  out-of-phase displacement of the Al/Si sublattice along z (B$_{1g}$ at
  $\Gamma$, B$_{2u}$ at A) becomes unstable around k$_z$= $\pm\pi/c$ for $x <
  0.2$ (imaginary frequencies are shown as negative).}
\label{fig:phonons_vs_x}
\end{figure}

In Fig.~\ref{fig:phonons_vs_x} we plot, as a function of $x$, the CaAl$_{2-x}$Si$_x$ phonon frequencies of the B$_{1g}$ branch calculated at $\Gamma$, A and L (purely imaginary frequencies are shown as negative values). 
At $x < 0.2$ there is a large jump in the phonon frequencies of the B$_{1g}$
branch at A and L, which become
imaginary (shown as negative), while the zone-center phonons are still all stable.
The entire CaAl$_2$  ($x\!=\!0$) phonon spectrum, shown in the lower panel of
Fig.~\ref{fig:CaAl2},
 reveals that the compound 
is dinamically unstable against lattice distorsions with $q_z\!=\!\pm\pi/c$
and eigenvectors corresponding to the $B_{1g}$ optical  
and $A_{2u}$ acoustical branches, both implying an out-of-plane displacement
of the Al - Si atoms.

\begin{figure}[h!tbp]
\begin{center}
\includegraphics[width=8cm]{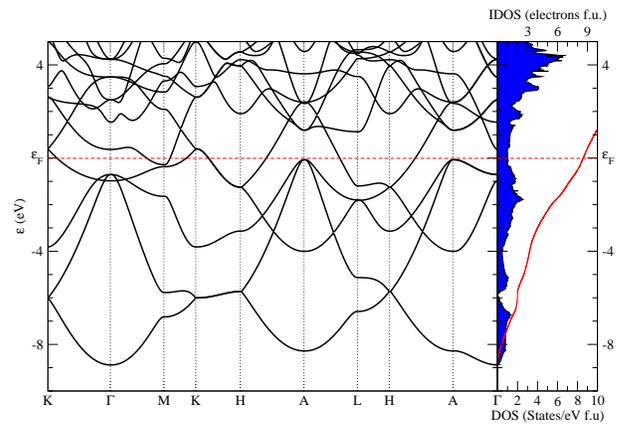}
\end{center}
\caption{Energy bands (black), electronic DOS (blue) and integrated DOS (solid red) at $x=$0.6, i.e., the lowest Si concentration for which CaAlSi is observed in the C$_{32}$ structure.}
\label{fig:bande_x0.6}
\end{figure}

\begin{figure}[h!tbp]
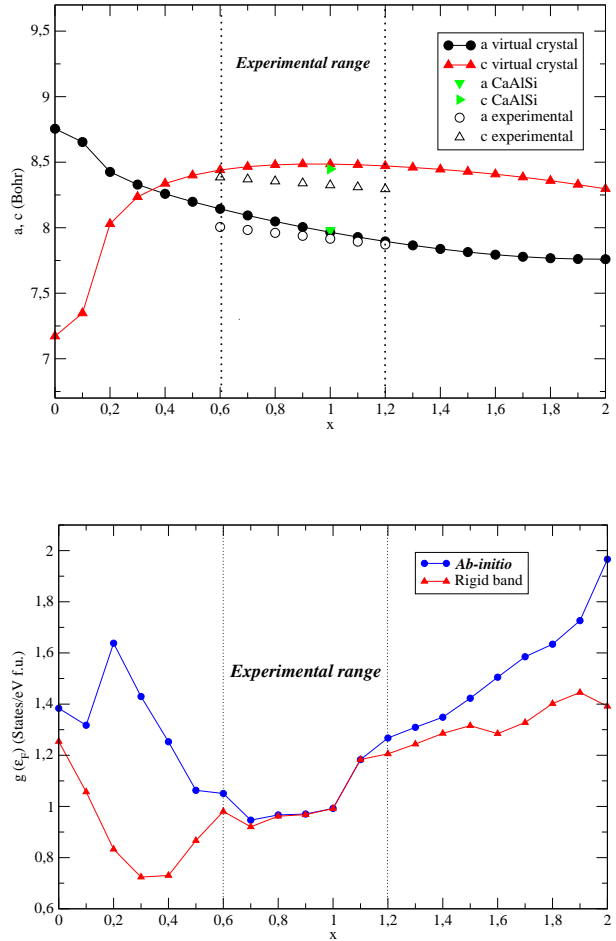

\vspace{1cm}
\begin{center}
\includegraphics[width=8cm]{lattice_vs_x}
\end{center}
\vspace{0.5cm}
\begin{center}
\includegraphics[width=8cm]{confronto_gef}
\end{center}
\caption{Upper panel: optimised lattice parameters for the C$_{32}$ crystalline structure of CaAl$_{2-x}$Si$_x$.
Lower panel: blue circles indicate the self-consistent density of states at the Fermi level $g(\epsilon_F)$ of CaAl$_{2-x}$Si$_x$ in the fully relaxed structure, while red triangles show $g(\epsilon_F)$ obtained with a rigid-band filling of Ca(AlSi) appropriate to the electron number $N_\text{e}(x)=8+x$ (see middle panel of Fig.~\ref{fig:bands_for_x=1}). }
\label{fig:lattice_parameters_rigidband}
\end{figure}

\begin{figure}[h!tbp]
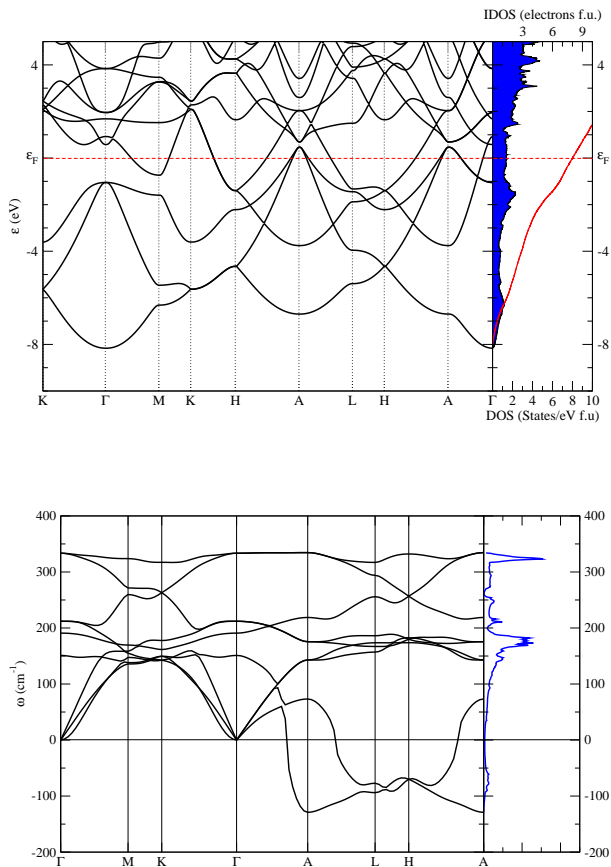

\begin{center}
\includegraphics[width=8cm]{0_0}
\end{center}
\vspace{0.4cm}
\begin{center}
\includegraphics[width=8cm]{ph_caal2_vc}
\end{center}
\caption{Upper panel: energy bands and DOS for CaAl$_2$ in the structurally relaxed  C$_{32}$ structure. Because of the very 
smalll $c/a$, the $\pi$ bands
acquire a  large dispersion in the $k_z$ direction, and are driven well above
$\epsilon_F$ in the $k_z$=0 plane.
Notice that the resulting Fermi surface (not shown) is still similar to that of Ca(AlSi). Lower panel: phonon dispersion and density of states corresponding to the same structure.}
\label{fig:CaAl2}
\end{figure}

The dramatic changes in the lattice parameters and phonon frequencies with
compositions $x < 0.6$ 
can be rationalised in terms of the electronic band structure.
We will start our discussion from
  the bands calculated at $x = 0.6$,  shown in Fig.~\ref{fig:bande_x0.6}. 
At this composition both the interlayer band and the $\pi$ bands cross the Fermi energy, while  
the $\sigma$ bands are still completely full, even though, around A, they are now very close to $\epsilon_F$. 
On these grounds one may think that, by further decreasing $x$ (i.e., by increasing the aluminium content), it would be possible to obtain $\sigma$ holes, as in MgB$_2$.
This simple hint proves, however, incorrect, since, as already mentioned (when discussing the lower panel of Fig.~\ref{fig:lattice_parameters_rigidband}), the rigid band model breaks down for $x <0.6$. 
We are going to make this statement more precise. 
Our first-principles calculations have revealed that the $\sigma$, $\pi$ and interlayer bands are no longer rigid when the aluminium content is 
increased above $70\%$ (i.e. for $x <0.6$).

The $\pi$ band decreases its binding energy on the $k_z\!\!=\!\!0$ plane, progressively losing electrons and getting 
emptier and emptier as $x$ is reduced, while the interlayer and the $\sigma$ bands remain partially filled to conserve the total 
number of electrons per unit cell; but then,
as $x$ further decreases, the crystal structure, and therefore the bands, 
substantially rearrange.
In a rigid band picture, based on the $x\!=\!1$ DOS, 
the interlayer band would become completely empty at
$x\!\!=\!\!0.5$ (see middle panel of Fig.\ref{fig:bands_for_x=1}); but,
precisely around this composition, the rigid-band picture starts to break
down, and the behaviour of the $c/a$ ratio starts to change in such a way 
as to keep the interlayer band always at $\epsilon_F$.

When, at $x =0.2$, the interlayer band is almost empty,
our self-consistent calculations
predict a further, abrupt change in $c/a$. This sudden relaxation
is related to a major rearrangement of all bands, whose main result is that the
interlayer band stays at the Fermi level, while the
$\pi$ (bonding) band becomes completely empty in the $k_z\!=\!0$ plane.

The complete emptying of the $\pi$ band in the ``upper half" of the Brillouin
zone 
($k_z\simeq \pm\pi/c$, $A-L-H-A$ path)
further weakens the bonds in the vertical direction, and this explains
the appearance of imaginary frequencies in the out-of-plane phonon branches of the $C_{32}$ phase of CaAl$_{2-x}$Si$_{x}$ for $x<0.2$. 
The displacement corresponding to these phonons is compatible with the stable
structure of CaAl$_2$ at ambient condition, the MgCu$_2$ Laves phase, in which
the Al atoms sit on triangular layers intercalated by mixed Ca--Al planes. 
In this structure, the main bonding force is provided by $p-d$ bonds between Ca and
Al, while in the C$_{32}$ 
structure the in-plane $sp^2$ bonding is important.
Interestingly, experiments have shown\cite{LAVES1} that the Laves phase  is stable only up to $x < 0.2$, where we find the occurrence of a lattice instability for the C$_{32}$ phase. A thorough investigation of the whole phase diagram and transition goes, however, beyond the aim of the present study.\cite{FN3}

In the upper panel of Fig.~\ref{fig:CaAl2} we display the electronic bands of CaAl$_2$ corresponding to the optimised $C_{32}$ structure. Because of the large modifications in the lattice
parameters, these bands bare very little resemblance to those of Ca(AlSi), and
even less to those of MgB$_2$. 
What seemed to be a good candidate for a compound with properties similar to
MgB$_2$ turns out to be a very poor 
imitation, once the lattice parameters are optimised. 
In particular, even though there is a tiny amount of $\sigma$ holes around the
A point --the fingerprint 
and main ingredient for superconductivity of MgB$_2$--, 
in CaAl$_2$ most of the DOS at $\epsilon_F$ comes from the $\pi^*$ bands  
and from the the interlayer band (which in MgB$_2$ is, instead, above  
$\epsilon_F$, and thus completely empty). Worst of all, as we already saw, 
CaAl$_2$ is dynamically unstable in the C$_{32}$ structure.

\begin{figure}[h!tbp]
\vspace{0.4cm}
\begin{center}
\includegraphics[width=8cm]{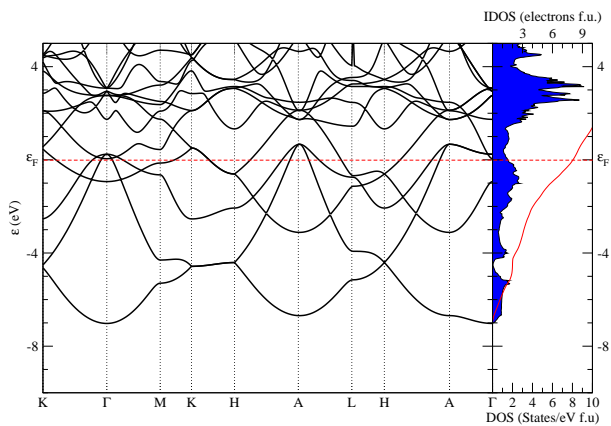}
\end{center}
\caption{Band structure of CaAl$_2$ in the C$_{32}$ structure with the 
same $c/a$ ratio as MgB$_2$ ($c/a\!=\!1.142$). In this case there are $\sigma$
holes at the Fermi level both at A and $\Gamma$
 and all the phonon frequencies (not shown)
are real.}
\label{fig:CAAL2_MGB2}
\end{figure}

We wish to point out that the differences in the band structure and the
dynamical instability are both a consequence of the $c/a$ reduction: 
if we not only constrain the compound to the C$_{32}$ structure, but also keep 
the $c/a$ ratio at the same value as MgB$_2$ (rather than being optimised, as done until now),
 then we find $\sigma$ holes at the Fermi level (see fig. \ref{fig:CAAL2_MGB2})
 and a dynamically stable compound: all the calculated phonon frequencies, not
 shown, are now real.\cite{FN4}

We finally observe that the abrupt $c/a$ reduction (when the crystal is, as in
our calculations, artificially forced to maintain the C$_{32}$ structure)
amounts to a Fermi-level pinning 
of the bottom of the interlayer band. Since, as mentioned, the bottom of the
interlayer band has a prevailing
 Ca-$d$ character, this observation in turn  suggests that the mechanism which
 makes CaAl$_2$ so different 
from MgB$_2$ and drives its instability towards the Laves phase is, in fact, 
the pinning at the Fermi level of the calcium $d$-states.

\subsubsection*{High silicon concentration, $x > 1.2$}

While in the limit of high aluminium concentration ($x\!<\!0.6$) we found
 several fingerprints and a plausible explanation for the experimental lattice
 instability, in the opposite limit (high silicon concentration, $x\!>\!1.2$)
 the only remarkable finding remains the deviation from the rigid-band
 behaviour, which, as $x$ increases, manifests itself in the slow descent
in energy of the lowest antibonding $\sigma$ band. This band, which at $x=1$
is $\sim 1$ eV above $E_F$ at $\Gamma$, starts filling up at $x=1.6$.

Apart from this deviation (which however is less dramatic than for $x\!<\!0.6$, see lower
panel of Fig.~\ref{fig:lattice_parameters_rigidband}),
neither the optimised structural parameters $a$ and $c$ (upper 
panel of Fig.~\ref{fig:lattice_parameters_rigidband}), nor the
phonon frequencies selected in Fig.~\ref{fig:phonons_vs_x},
nor, in fact, any frequency of the complete phonon spectrum
(not shown), seem to do anything special in the composition range
$1.2 \leq x \leq 2$. This suggests that (at zero temperature) the C$_{32}$ phase of CaSi$_2$ (i.e., $x\!=\!2$) represents a metastable structure, i.e., a local minimum of its total energy which is higher than the absolute  minimum. Such a guess is reinforced by the complexity of the unit cells of the actual stable phases of CaSi$_2$\cite{CASI2:Fahy},
which could easily imply sizable energy barriers separating them from the
 metastable C$_{32}$ phase.\cite{FN5}

Again, a thorough investigation of the whole phase diagram and transitions 
of CaSi$_2$ goes beyond the aim of the present study.

\section{Summary and Conclusions}
In this work we have reported first-principles calculations of the electronic
and vibrational properties of the ternary alloy 
CaAl$_{2-x}$Si$_x$ in the entire composition range $0 \leq x \leq 2$, discussing how electronic, structural and vibrational properties are related to each other. In this context an interlayer band, which is empty in MgB$_2$, and whose bottom has Ca-$d$ character in our compounds, plays a major role.

At $x\!\!=\!\!1$ the electronic band dispersion and Fermi surfaces, calculated considering
both an ordered phase with atomic in-plane ordering and a disordered VCA phase, 
confirms the suggestion of Mazin \emph{et al.}~\cite{MazinCaAlSi},
that the compound crystallises in a completely in-plane ordered phase,
if synthesised in presence of equal amounts of Al and Si.
Our calculation of the electron-phonon coupling confirms
that the soft B$_{1g}$ phonon branch plays a major role in the superconductivity of CaAlSi, and shows that this mode couples the interlayer and $\pi^*$ bands. At variance with a previous full-potential calculation, we do not find any instability in the phonon spectrum of CaAlSi.

At  $x \neq 1$, in the experimentally accessible composition range,
the CaAl$_{2-x}$Si$_x$ electronic bands are accurately described by a rigid-band model; in this range
$g(\epsilon_F)$ monotonically increases with $x$, so that 
the measured non-monotonic variations in the superconducting properties cannot be associated with changes in the density of states at the Fermi level. These findings provide a useful reference and a strong motivation for a future, complete study of the electron-phonon interaction as a function of $x$.

The rigid-band model breaks down for $x<0.6$, where the increasing aluminium content mainly leads to the emptying of the $\pi$ bonding band, which finally drives the compound unstable. The occurrence of such a dynamical instability is related to an abrupt change in the $c/a$ ratio and correlates with the persistence of the interlayer band at the Fermi level, related to the pinning of the Ca-$d$ states.
Therefore, although isostructural and isoelectronic with  MgB$_2$, 
the C$_{32}$ phase of CaAl$_2$ is unstable and doesn't share strong
similarities with its electronic bands.

The main result of this work is that the interlayer band, empty in MgB$_2$,
and the ``out-of-plane" phonons 
(optical $B_{1g}$ and acoustical $A_{2u}$ at $\Gamma$), irrelevant for
superconductivity in MgB$_2$, 
play a major role on the stability and superconductivity of C$_{32}$
intermetallic compounds, 
once the structural parameters $c$ and $a$ assume appropriate values. This observation goes beyond the CaAlSi family and represents the starting point for further studies of a wider class of hexagonal, graphite-like compounds. \cite{oka}

\section{Acknowledgments}
We are very grateful to Ole K. Andersen for useful conversations and to Jens
Kortus 
for a critical reading of this manuscript. One of us (GBB) gratefully
acknowledges 
partial financial support from MIUR (the Italian Ministry for Education,
University and Research) 
through COFIN2003.


\begin{thebibliography}{99}

\bibitem{mgb2discovery}
{J. Nagamatsu, N. Nakagawa, T. Muranaka, Y. Zenitani and J. Akimitsu}.
\newblock {\em Nature (London)}, \textbf{410}:63, (2001).

\bibitem{c32type}
{P. Villars and L.D. Calvert}.
\newblock {\em {\textit{Pearson's Handbook of Crystallographic Data for
  Intermetallic Phases}, 2nd ed.}}
\newblock {ASM International, Materials Park, OH,}, (1991).

\bibitem{CaAlSiordered}
{B. Lorenz, J. Lenzi, J. Cmaidalka, R. L. Meng, Y. Y. Sun, Y. Y. Xue and C. W.
  Chu}.
\newblock {\em Physica C}, \textbf{383},191, (2002).

\bibitem{CaAlSi_pressione_exp}
{B. Lorenz, J. Cmaidalka, R. L. Meng and C. W. Chu}.
\newblock {\em {Phys. Rev. B}}, \textbf{68},14512, (2003).

\bibitem{unconventional_SC_CaAlSi}
{S. Kuroiea, H. Takagiwa, M. Yamazawa and J. Akimitsu}.
\newblock {\em {cond-mat}}, \textbf{0402483}, (2004).

\bibitem{CaAlSi-Fermi}
{S. Tsuda, T. Yokoya, S. Shin, M. Imai, I. Hase}.
\newblock {\em Phys. Rev. B}, \textbf{69}, 100506, (2004).

\bibitem{SheinCaAlSi}
{I.R. Shein, N.I. Medvedeva, A.L. Ivanovskii}.
\newblock {\em J. Phys.: Condens. Matter}, \textbf{15}, L541, (2003).

\bibitem{MazinCaAlSi}
{I.I. Mazin, D.A. Papaconstantopoulos}.
\newblock {\em Phys. Rev. B}, \textbf{69}, 1805, (2004).

\bibitem{CaAlSifononi-cinesi}
{G.Q. Huang, L.F. Chen, M. Liu, D.Y. Xing}.
\newblock {\em Phys. Rev. B}, \textbf{69}, 064509, (2004).

\bibitem{CaAlSi_pressure_effects}
{G. Q. Huang, L. F. Chen, M. Liu and D. Y. Xing}.
\newblock {\em {Phys. Rev. B}}, \textbf{71}, 172506, (2005).

\bibitem{FN1}
In reality, CaAl$_2$ crystallises in the so-called MgCu$_2$ Laves phase, and,
  when doped with Li or Mg, in the MgNi$_2$ and MgZn$_2$ phases.~\cite{LAVES1,
  LAVES2, LAVES3}; while theoretical and experimental work exists on these
  transitions, the effect of Si doping is still unexplored.

\bibitem{LAVES1}
H.Tanaka, H.Takeshita, N.~Kuriyama, T.~Sakai, I.~Uehara, D.~Noreus,
  A.~Z\"uttel, L.~Schlapbach, and S.Suda.
\newblock {\em IEA Task 12: Metal Hydrides and Carbon for Hydrogen Storage},
  page~23, 2001.

\bibitem{LAVES2}
S.~Amerioun, S.~I. Simak, and U.~Haussermann.
\newblock {\em Inorg. Chem.}, \textbf{42}, 1467, (2003).

\bibitem{LAVES3}
S.~Amerioun, T.~Yokosawa, S.~Livin, and U.~Haussermann.
\newblock {\em Inorg. Chem.}, \textbf{43}, 4751, (2004).

\bibitem{MGBC}
C.~Petrovic R.A.~Ribeiro, S.L.~Bud'ko and P.C. Canfield.
\newblock {\em Physica C}, \textbf{384}, 227, (2003).

\bibitem{DFT:KS}
W.~Kohn and L.~J. Sham.
\newblock {\em Phys. Rev.}, \textbf{40}, A1133, (1965).

\bibitem{DFT:HK}
P.~Hohenberg and W.~Kohn.
\newblock {\em Phys. Rev.}, \textbf{136}, B864, (1964).

\bibitem{DFT:BaroniRMP}
S.~Baroni, S.~de~Gironcoli, A.~Dal Corso, and P.~Giannozzi.
\newblock {\em Rev. Mod. Phys.}, \textbf{73}, 515, (2001).

\bibitem{ABINIT}
{The ABINIT code is a common project of the Universit\'e Catholique de Louvain,
  Corning Incorporated, and other contributors. URL http://www.abinit.org}.

\bibitem{PWscf}
{S. Baroni, A. Dal Corso, S. de Gironcoli, P. Giannozzi, C. Cavazzoni, G.
  Ballabio, S. Scandolo, G. Chiarotti, P. Focher, A. Pasquarello, et al. URL
  http://www.pwscf.org}.

\bibitem{Troullier-Martins}
{N. Troullier, J. L. Martins}.
\newblock {\em Phys. Rev. B}, \textbf{43}, 1993, (1991).

\bibitem{fhi98pp}
{M. Fuchs, M. Scheffler}.
\newblock {\em {Comput. Phys. Commun.}}, \textbf{119}, 67, (1999).
\newblock {URL http://www.fhi-berlin.mpg.de/th/fhi98md/fhi98PP/}.

\bibitem{PBEfunctional}
{J.P. Perdew, K. Burke and M. Ernzerhof}.
\newblock {\em Phys. Rev. Lett.}, \textbf{77}, 3865, (1996).

\bibitem{MPgrid}
{H.J. Monkhorst, J.D. Pack}.
\newblock {\em Phys. Rev. B}, \textbf{13}, 5188, (1976).

\bibitem{tetrahedron}
{P.E. Bl\"ochl, O. Jepsen, O. K. Andersen}.
\newblock {\em Phys. Rev. B}, \textbf{49}, 16223, (1994).

\bibitem{Coldsmearing}
{N. Marzari, D. Vanderbilt, A. De Vita, M. C. Payne}.
\newblock {\em Phys. Rev. Lett., \textbf{82}, 3296, (1999).}
\bibitem{VCA}
{L. Nordheim}.
\newblock {\em Ann. Phys. Leipzig}, \textbf{9}, 607, (1931).

\bibitem{CaAlSidisorderd}
{M. Imai, K. Nishida, T. Kimura and H. Abe}.
\newblock {\em Appl. Phys. Lett.}, \textbf{80}, 1019, (2001).

\bibitem{mg:kortus1}
J.~Kortus, I.~I. Mazin, K.~D. Belashchenko, V.~P. Antropov, and L.~L. Boyer.
\newblock {\em Phys. Rev. Lett.}, \textbf{86}, 4656, (2001).

\bibitem{mg:pickett}
J.~M. An and W.~E. Pickett.
\newblock {\em Phys. Rev. Lett.}, \textbf{86}, 4366, (2001).

\bibitem{posternack}
{M. Posternak, A. Baldereschi, A. J. Freeman, E. Wimmer, and M. Weinert}.
\newblock {\em {Phys. Rev. Lett.}}, \textbf{50}, 761, (1983).

\bibitem{mythesis}
{M. Giantomassi}.
\newblock {Elettroni, fononi e propriet\`a strutturali del composto
  CaAl$_{2-x}$Si$_x$}, (2005).
\newblock {Master Thesis, unpublished}.

\bibitem{mg:kong}
Y.~Kong, O.~V. Dolgov, O.~Jepsen, and O.~K. Andersen.
\newblock {\em Phys. Rev. B}, \textbf{64}, 020501, (2001).

\bibitem{mg:satta}
G.~Satta, G.~Profeta, F.~Bernardini, A.~Continenza, and S.~Massidda.
\newblock {\em Phys. Rev. B}, \textbf{64}, 104507, (2001).



\bibitem{MAURI_GRAF}
S.~Piscanec, M.~Lazzeri, Francesco Mauri, A.~C. Ferrari, and J.~Robertson.
\newblock {\em Phys. Rev. Lett.}, \textbf{93}, 185503, (2004).

\bibitem{pippo3}
Frozen-phonon calculations , not shown, confirm that the interlayer intraband
scattering is also negligible.

\bibitem{FN2}
In some cases, when there is a clear experimental hint for the stability of a
  certain compound in a given structure, and when the instability is limited to
  a few {\bf q}-points, it is possible to use a generalized approach to include
  the contribution of phonons with small, imaginary frequencies to the total
  electron-phonon coupling. We think that in our case this approach cannot be
  followed: here a whole branch has imaginary frequencies, which simply tells
  that our fictitious $\Box^{2+}$AlSi is not stable.

\bibitem{FN3}
In the MgCu$_{2}$ Laves phase (space group No. 227)~\cite{c32type} the Ca atoms
  sit on (8a) and Al on (16d) sites, with a lattice parameter of 8.04 \AA. We
  checked that the total energy of CaAl$_2$, in the optimised Laves phase
  structure ($a=8.06 \AA$), is lower than in the AlB$_2$ structure by 0.2
  eV/f.u.

\bibitem{FN4}
In other words, this hypotetical C$_{32}$ is stable against lattice vibrations
  at fixed $c$ and $a$, although obviously unstable against structural
  relaxation of $c$ and $a$.

\bibitem{CASI2:Fahy}
S.~Fahy and D.R. Hamann.
\newblock {\em Phys. Rev. B}, \textbf{41}, 7587, (1990).

\bibitem{FN5}
At P=16 GPa CaSi$_2$ crystallises in a slightly distorted C$_{32}$ phase
  \cite{CASI2_exp}, which has been studied by \emph{Satta et
  al.}\cite{mg:satta}.


\bibitem{CASI2_exp}
P.~Bordet, M.~Affronte, S.~Sanfilippo, M.~Nunez-Regueiro, O.~Laborde, G.~L.
  Olcese, and A.~Palenzona.
\newblock {\em Phys. Rev. B}, \textbf{62}, 11392, (2000).

\bibitem{oka}
L.~Boeri, M.~Giantomassi, G.B. Bachelet, and O.K. Andersen.
\newblock {\em Unpublished}.


\end{thebibliography}
\end{document}